\documentclass[a4paper,fleqn]{cas-dc}

\usepackage[numbers]{natbib}
\usepackage[linesnumbered,ruled,vlined]{algorithm2e}
\usepackage{pdfpages}
\def\tsc#1{\csdef{#1}{\textsc{\lowercase{#1}}\xspace}}
\tsc{WGM}
\tsc{QE}
\tsc{EP}
\tsc{PMS}
\tsc{BEC}
\tsc{DE}

\SetKwInput{KwInput}{Input}                
\SetKwInput{KwOutput}{Output}              

\begin{document}
\let\WriteBookmarks\relax
\def\floatpagepagefraction{1}
\def\textpagefraction{.001}

\shorttitle{A General Deep Learning framework for Neuron Instance Segmentation}

\shortauthors{Huaqian Wu et~al.}

\title [mode = title]{A General Deep Learning framework for Neuron Instance Segmentation based on Efficient UNet and Morphological Post-processing}                      

\author[1]{Huaqian Wu}
\author[1]{Nicolas Souedet}
\author[1]{Caroline Jan}
\author[2]{Cédric Clouchoux}
\author[1]{Thierry Delzescaux}[orcid=0000-0002-6527-7946]

\cormark[1]


\ead{thierry.delzescaux@cea.fr}

\address[1]{University of Paris-Saclay, CEA-CNRS-UMR 9199,
    MIRCen, 
    Fontenay-aux-Roses,
    France}

\credit{Data curation, Writing - Original draft preparation}


\address[2]{WITSEE,
    Paris,
    France}

\cortext[cor1]{Corresponding author}

\begin{abstract}
  Recent studies have demonstrated the superiority of deep learning in medical image analysis, especially in cell instance segmentation, a fundamental step for many biological studies. However, the excellent performance of the neural networks requires training on large, unbiased dataset and annotations, which is labor-intensive and expertise-demanding. This paper presents an end-to-end framework to automatically detect and segment NeuN stained neuronal cells on histological images using only point annotations. Unlike traditional nuclei segmentation with point annotation, we propose using point annotation and binary segmentation to synthesize pixel-level annotations. The synthetic masks are used as the ground truth to train the neural network, a U-Net-like architecture with a state-of-the-art network, EfficientNet, as the encoder. Validation results show the superiority of our model compared to other recent methods. In addition, we investigated multiple post-processing schemes and proposed an original strategy to convert the probability map into segmented instances using ultimate erosion and dynamic reconstruction. This approach is easy to configure and outperforms other classical post-processing techniques. This work aims to develop a robust and efficient framework for analyzing neurons using optical microscopic data, which can be used in preclinical biological studies and, more specifically, in the context of neurodegenerative diseases.
\end{abstract}

\begin{highlights}
\item A mask-synthesis pipeline to generate pixel-level labels using only point annotations.
\item A thorough comparison of accuracy and computation cost of neural networks. 
\item A novel post-processing strategy to refine the instance segmentation.
\end{highlights}


\maketitle

\section{Introduction}

Advances in microscopy allow scanning whole slide images, capturing details at the cellular level and revealing the complexity of brain structures. It provides the opportunity to quantitatively analyze cell populations, morphology and distribution to answer biological questions. For example, the number, distribution \cite{karlsen2011total, hughes1987morphometric, thu2010cell} and morphometric information \cite{vicar2020quantitative} of neurons are important features in studying brain aging, including neurodegenerative diseases. A crucial prerequisite for such studies is cell instance segmentation, which plays a crucial role in digital pathology image analysis. Neuron segmentation is exceptionally challenging because the size, density, and intensity of neurons differ a lot from one anatomical region to another. Since manual identification of single cells is extremely laborious and time consuming, several automatic segmentation algorithms have been proposed: thresholding \cite{otsu1979threshold},  graph cut \cite{lou2012learning, he2015icut} and watershed \cite{cousty2008watershed,veta2011marker, veta2013automatic}. These methods need to be specifically adapted for different configurations (species, cell types, stainings). Furthermore, noise or other technical artifacts can easily influence the segmentation results. Under-segmentation and over-segmentation often occur when they deal with touching or overlapping cells like neurons. You et al. \cite{you2019automated} proposed a framework based on gaussian, min-max filter and region growing algorithm to deal with such data, but it is computationally expensive due to numerous iterations and performed poorly on light-stained regions.

Recently, deep learning (DL) has achieved remarkable progress in many fields \cite{lecun2015deep}, especially in medical image analysis. Neural networks have been successfully applied to detect abnormal signals \cite{zhou2021recognition}, segment lesion areas \cite{wang2020multi, tang2021construction, he2022image, zhang2021adoption} for clinical diagnosis. DL-based methods have also shown superiority in cell segmentation competitions \cite{caicedo2019nucleus,kumar2019multi}, achieving better segmentations with stronger robustness than traditional algorithms. Naylor et al. \cite{naylor2018segmentation} addressed this problem as a regression task of estimating the nuclei distance map. A more common strategy is to address this problem as a semantic segmentation task, such as the pixel-wise binary classification of cells and background \cite{sirinukunwattana2016locality}, or more recently, the ternary classification of the interior of cells, background and cell boundaries \cite{caicedo2019nucleus, cui2019deep,chen2017dcan}. Most convolutional neural networks (CNNs) such as AlexNet \cite{krizhevsky2017imagenet}, VGGNet \cite{simonyan2014very} and ResNet \cite{he2016deep} learn representations by gradually reducing the size of feature maps, but the high-resolution features are lost during this process. These networks are not suitable for pixel-wise tasks like cell segmentation. Several networks add a resolution-recover process to address segmentation problems. For example, SegNet \cite{badrinarayanan2017segnet} and DeconvNet \cite{noh2015learning} use unpooling and deconvolution layers to recover original resolution; U-Net \cite{ronneberger2015u}, a breakthrough in the field of medical image segmentation, with skip connections concatenate the high resolution features of the encoder path to the upsampled output of the decoder path. Our previous work \cite{wu2021evaluation} evaluated an ensemble model of eight U-Net-like neural networks with different backbone CNNs. Wang et al. proposed HRNet \cite{wang2020deep}, which maintains high-resolution representations and assembles features from multi-resolution streams. It outperformed other state-of-the-art networks on several tasks of semantic segmentation, object detection and instance segmentation.  

The good performance of CNNs relies on large datasets and the quality of pixel-level annotations, which are tedious and labor-intensive to be carried out manually. To facilitate the labeling process, researchers investigated several weakly-supervised methods using point annotations: Qu et al. \cite{qu2020weakly} trained a CNN to predict the cell center location, and generated pixel-level labels using Voronoi transformation and k-means clustering. Based on a similar strategy, Chamanzar et al. \cite{chamanzar2020weakly} proposed a multi-task learning integrating repel encoding to enhance segmentation performance. However, \cite{qu2020weakly, chamanzar2020weakly} are not straightforward and involve multiple networks or branches to achieve the final segmentation. 

Moreover, CNNs without post-processing often failed to handle touching objects \cite{naylor2018segmentation}. Researchers mainly focus on improving the performance of CNNs, while the post-processing part is generally not explicitly described, although it is a critical step to obtain good segmentations. Applying watershed segmentation (WS) on the cell probability map derived from DL is the most common way \cite{kumar2019multi, xie2020integrating}. Graph partition \cite{song2015accurate} and distance transform \cite{xing2015automatic} are also popular techniques. The winning method of \cite{caicedo2019nucleus} proposed a more tricky technique: a regression model was trained to predict the intersection-over-union (IoU) for cell candidates produced by applying different thresholds on the probability map. With this method, only the candidate with the highest IoU was preserved for each object. One drawback of these methods is that well-configured parameters are required to ensure the performance. Thus, this requires redesign of the parameter settings for applications on novel data. Generic post-processing methods for cell instance segmentation are scarce and worth investigating.

In this paper, we propose an end-to-end framework based on CNNs for neuron instance segmentation, with the following contributions: 1) we establish and validate a semi-automated pixel-level-mask synthesis pipeline using only point annotations and Random Forest (RF) binary segmentations. This approach allows to generate a large labeled dataset with minimal manual cost, 2) inspired by the instance segmentation challenge \cite{van2021spacenet}, we integrate EfficientNet-B5 \cite{tan2019efficientnet} into a U-Net-like encoder-decoder architecture, this new model is spatially and semantically precise and 3) we propose a novel strategy for post-processing probability map based on ultimate erosion, dynamic reconstruction and WS. Our framework is generic and easy to configure: the mask synthesis pipeline is independent of the neural network, the synthetic masks can be easily derived to validate other supervised methods. The only parameter needed at the post-processing stage is the size of the structuring element, which equals the size of the smallest cell in the dataset. We compared our network to various state-of-the-art CNNs methods \cite{you2019automated, cui2019deep, wu2021evaluation, wang2020deep}, as well as the proposed post-processing to other classical approaches. The findings demonstrate the superiority of our framework in terms of accuracy and efficiency, allowing us to develop a powerful tool for evaluating neurons in preclinical research and neurodegenerative diseases.  

\section{Material and methods}
\subsection{Dataset}
The data of this study are a set of two-dimensional (2D) light microscopy images. A representative histological section with a thickness of 40 $\mu m$ was produced from a healthy 9-year-old male macaque brain, stained by immunohistochemistry using the neuronal nuclei (NeuN) antibody, and scanned by an AxioScan.Z1 (Zeiss) with the in-plane resolution of 0.22 $\mu m$/pixel ($\times$20 magnification). The animal study was reviewed and approved by the Comité d’éthique approved by the MESR belonging to the EU: CETEA DSV – Comité n◦44. Thirty images of 5000 $\times$ 5000 pixels were selected to represent the heterogeneity of neuron distribution in the central anatomical regions, including cortex, hippocampus, thalamus, subiculum, etc (see Supplementary Fig. \ref{fig1}). These images showed a large diversity in terms of neuron size, shape, contrast and density, with both sparse (e.g. caudate and thalamus) and highly aggregated regions (e.g. cortex and hippocampus), as illustrated in Fig. \ref{fig1}. The pie chart presents the dataset composition in terms of anatomical regions, with about 67\% of images coming from the cortex and the hippocampus since the cortex is the largest brain structure (76\%) \cite{anand2012hippocampus} and hippocampal atrophy is linked to several neurodegenerative diseases \cite{van2008visualizing}. 

\begin{figure*}
\centering
\includegraphics[width=0.9\linewidth]{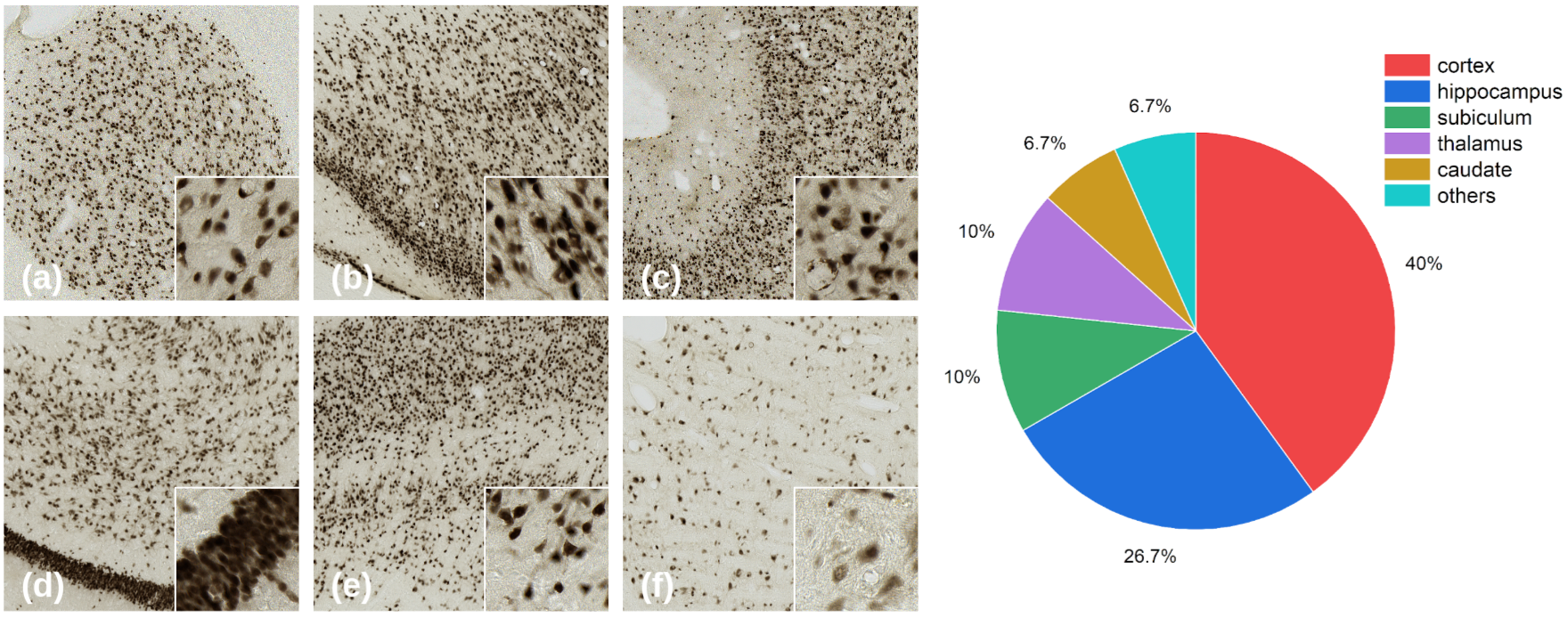}
\caption{Dataset examples with magnifications. (a) caudate, (b,c) cortex, (d) hippocampus, (e) subiculum and (f) thalamus.}
\label{fig1}
\end{figure*}

To verify the representativeness of our test set, we extracted features of the dataset using a ResNet101 model \cite{he2016deep} pre-trained on ImageNet and embedded the high-dimensional features into 2D space with t-distributed stochastic neighbor embedding (t-SNE) projection \cite{van2008visualizing}.

All images were divided into two subsets: 24 images for the training set and 6 images for the test set. The test set contains the following regions: caudate, cortex ($\times$ 2), hippocampus, subiculum and thalamus. Training images were cropped into 11k patches of 224 $\times$ 224 pixels for which ¼ of them were used to validate our neural networks at the end of each training epoch. To prevent overfitting and increase the robustness of the model, we applied data augmentation, including random rotation, vertical and horizontal flipping, RGB channel shuffling, color inversion, etc. The training set was expanded to 6 times the original size.

The size of images in the test set is 5000 $\times$ 5000 pixels, which requires more memory than the GPU RAM. They were firstly cropped into smaller patches (1344 $\times$ 1344 pixels, can be adjusted according to the GPU RAM) with an overlap (120 pixels, ~10\%) in both vertical and horizontal directions. The size of patches was constrained by the GPU memory (16 GB). A weighted map \cite{cui2019deep} was applied to the probability map of each patch to reduce the impact of inaccurate prediction at the border area. The weighted probability maps were then seamlessly assembled to reconstruct the probability map of the original large-scale image.  

In order to assess the generalizability of the proposed method, we added a test set that is independent of our training set. As shown in Supplementary Fig. \ref{fig2}, this dataset contains four cortex images of 1024 $\times$ 1024 pixels from various animal subjects, including two macaques, a microcebus \cite{gary2019encephalopathy}, and a mouse. Manual point annotation is conducted to evaluate the object-level segmentation of the proposed method. This study focuses on the cortex, which is not only the largest brain structure but also the region of greatest interest to neuroscientists. Compared to the training set (see Fig. \ref{fig1}), the additional macaque images are less brown and have greater contrast, the microcebus image is rosy brown, and the mouse image is gray with lighter stain intensity. 

\subsection{Pixel-level mask synthesis}
\begin{figure*}
\centering
\includegraphics[width=0.9\linewidth]{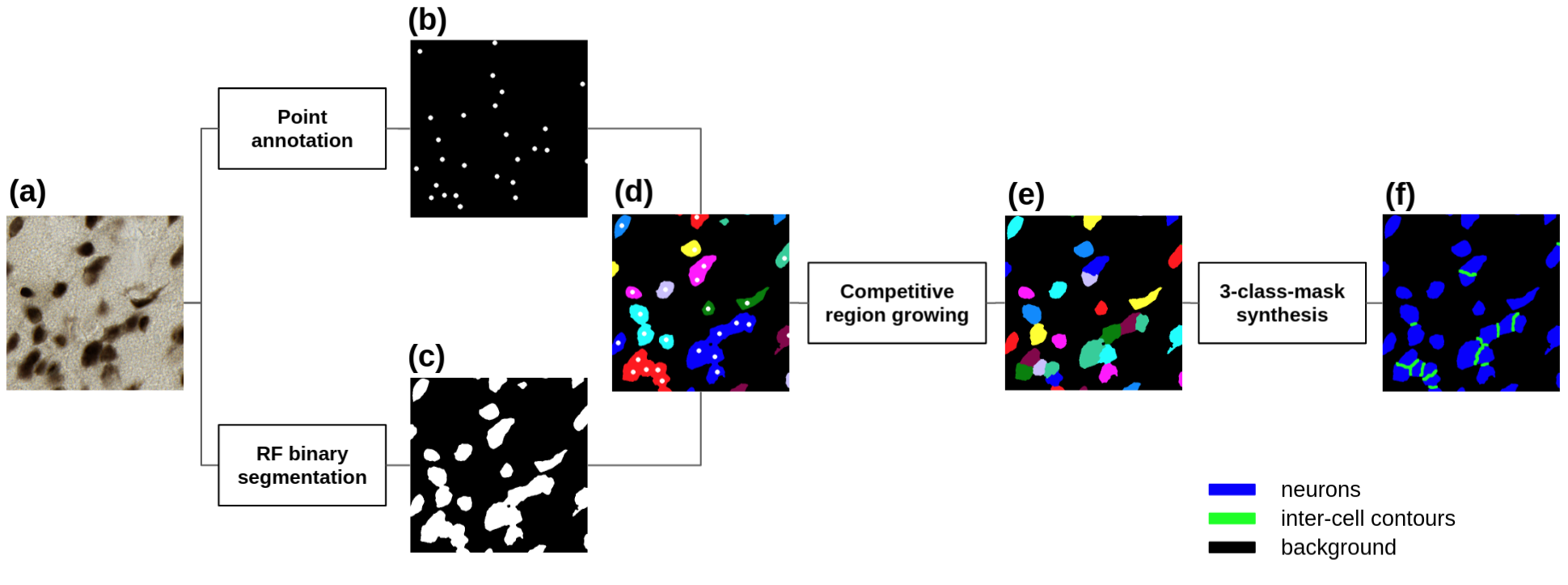}
\caption{ Pixel-level mask synthesis. (a) original image, (b) manual point annotations, (c) RF binary segmentation, (d) fusion of colored connected components and (c), (e) labeled image produced by region growing and (f) final three-class-masks, blue: neurons, green: inter-cell contours and black: tissue.}
\label{fig2}
\end{figure*}

Traditional nuclei segmentation methods based on point annotations involve center detection and pixel-level label extraction. In this study, we addressed neuron instance segmentation as a semantic segmentation task, which is to classify each pixel in the image into three following classes: neurons, background and contours of touching neurons. However, this strategy required instance annotations at the pixel level, which would have been extremely labor-intensive to achieve. We designed a pipeline to synthesize pixel-level masks to alleviate the manual labeling effort. Our strategy is shown in Fig. \ref{fig2}, which consists of two stages: the first stage is to segment neurons from the tissue and perform point annotation, and the second stage is to further separate each neuron instance under the guidance of point annotations for initialization and the constraint of the binary segmentation. Fig. \ref{fig2} (a) shows a NeuN image, where the neuron centroids were annotated manually by a disk with a radius of 5 pixels, as shown in Fig. \ref{fig2} (b). Fig. \ref{fig2} (c) presents the binary segmentation of neurons and the tissue, which is generated automatically with a RF model. As described in the previous work \cite{you2019automated}, the RF model contains 100 decision trees. It is trained using an optimized subset of automatically selected features on a small binary segmentation dataset of 100 images of 512 $\times$ 512 pixels \cite{bouvier2021reduced}. A CNN may be less effective with a small training dataset of this size. Fig. \ref{fig2} (d) shows the connected components of the binary segmentation superimposed with point annotations. Multiple centroids can be found inside one component since the binary segmentation could not identify individual neurons. Thus, the second phase involved a competitive region growing algorithm separating touching neurons, using centroids as the seed to initialize the growing process, and the binary segmentation of RF to constrain the expansion. Fig. \ref{fig2} (e) shows the results of the region growing, where neuron instances are separated and assigned with a unique label. The generated annotations were used as ground truth since the process was guided by the point annotation. Morphological operations were applied to instance annotations to obtain masks of three classes: in addition to the neuron and tissue classes, the pixels that connect different labels were automatically identified as the third class of inter-cell contour to separate touching or overlapping neurons \cite{wu2021evaluation}. The inter-cell contour has a constant thickness of 4 pixels, which was defined empirically. Fig. \ref{fig2} (f) shows the final result of the pipeline, which are synthetic masks of three classes that were used to train the segmentation neural networks. Although manual efforts are still required, such as the point annotations and the binary masks to train the RF, they are far less time-consuming and labor-intensive than massive instance annotations required to train a CNN for instance segmentation.

It is worth noting that the objective was to generate pixel-level annotations to complete the dataset needed to train the segmentation neural networks. The use of this pipeline is finished once the annotations are synthesized, but the same strategy can be used to annotate other datasets. 

To validate the synthetic mask generation process, three experts performed manual segmentation on a small dataset containing five patches of 500 $\times$ 500 pixels, including caudate, cortex ($\times$2), hippocampus and subiculum. The average manual segmentation time for 5 images was 2.5 hours.

\subsection{Neural networks}
\begin{figure*}
\centering
\includegraphics[width=0.9\linewidth]{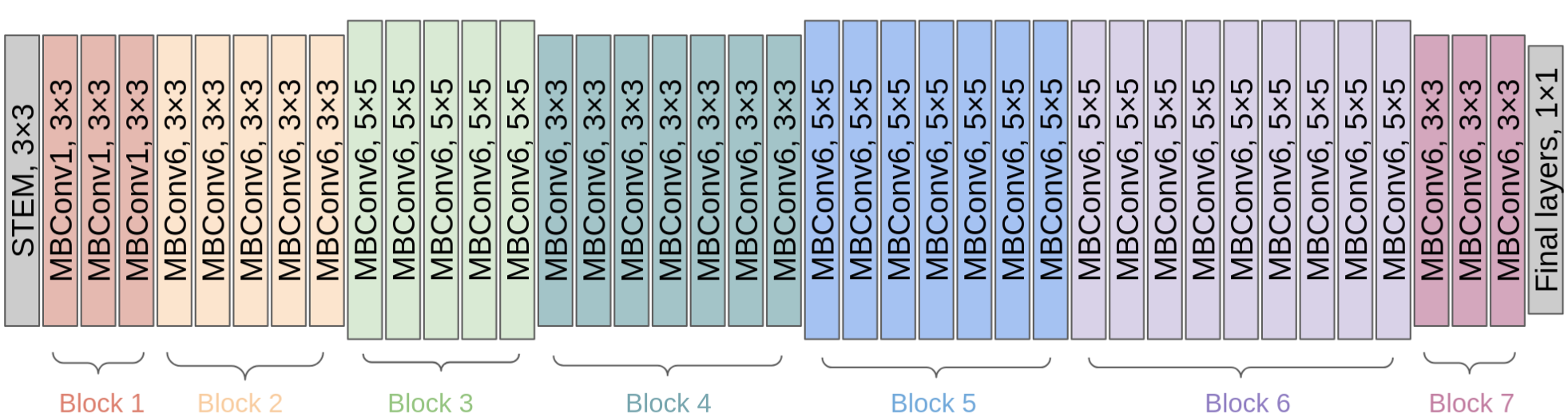}
\includegraphics[width=0.9\linewidth]{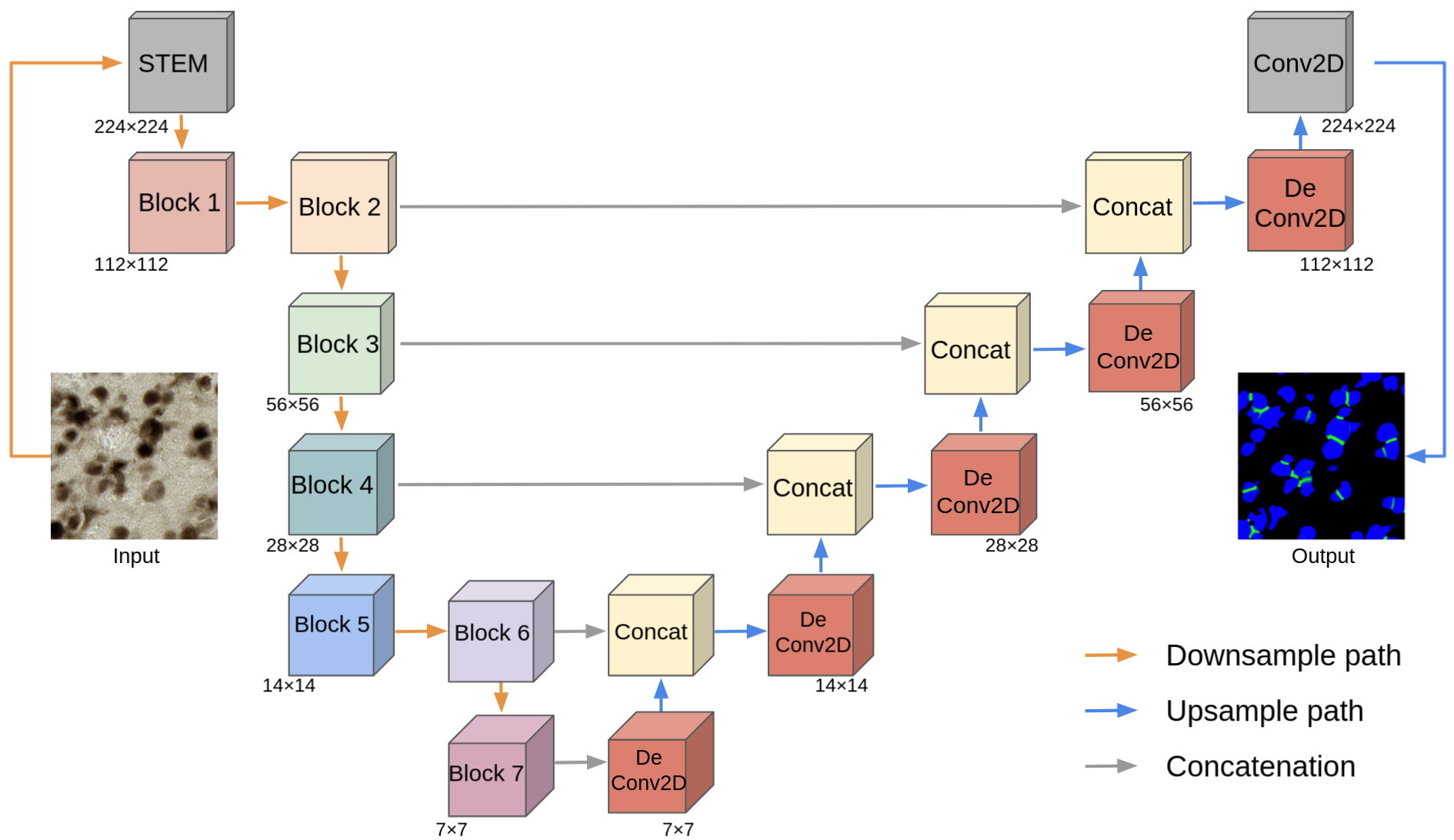}
\caption{Top: structure of EfficientNet-B5, it consists of 7 building MBConv blocks (represented with different color) and bottom: structure of our neural network using EfficientNet-B5 as encoder, named EfficientUNet-B5, the encoder is concatenated with the decoder at five different resolution (Block 2, Block 3,  Block 4, Block 6 and Block 7).}
\label{fig3}
\end{figure*}

In our previous work \cite{wu2021evaluation}, we showed the efficiency of U-Net-like architecture for neuron instance segmentation. Hence, we decided to keep the same strategy with a more recent neural network as the backbone. The family of models called EfficientNets was proposed by Tan et al. \cite{tan2019efficientnet},  which showed superiority in accuracy and efficiency against previous CNNs. The baseline model, EfficientNet-B0, was generated with neural architecture search \cite{tan2019efficientnet}. Its main building block is mobile inverted bottleneck MBConv \cite{sandler2018mobilenetv2}. Scaling up one network dimension of width, depth and input image resolution can improve accuracy. In particular, compound scaling of three dimensions can provide a significant gain \cite{tan2019efficientnet}. This approach brought seven scaled-up versions, named EfficientNet-B1 to EfficientNet-B7. EfficientNet-B5 was  chosen in this work as the result of a trade-off between accuracy and training cost. Fig. \ref{fig3} (a) presents the architecture of EfficientNet-B5, consisting of stem layers, seven main building blocks of MBConv and final layers. The resolution of the feature map was reduced five times gradually (from 224 $\times$ 224 to 7 $\times$ 7 pixels) after the stem layers, block 2, block 3, block 4 and block 6, respectively. Based on EfficientNet-B5 as the encoder, we gradually recovered the original high resolution through the decoder path, which consists of deconvolution and convolution layers. Fig. \ref{fig3} (b) shows the structure of our network, named EfficientUNet-B5, skip connections concatenate encoder and decoder at five different resolutions.  

\subsection{Loss function}
Since the neuron instance segmentation was addressed as a semantic segmentation of three classes, we used the compound loss of categorical cross-entropy (CE), two soft dice losses ( $D_{neuron}$ and $D_{contour}$) \cite{wu2021evaluation,zhou2019unet++} for neuron class and inter-cell contour class respectively to train the network. The global loss function L is defined as:
\begin{equation}
L = 0.5CE + 0.3D_{neuron} + 0.2D_{contour}
\end{equation}

\begin{equation}
  CE = -\frac{1}{nc} \sum_{i,j} \sum_{k}^{c} t_{i,j,k} \log(p(i,j,k))
\end{equation}

\begin{equation}
 D_k = 1-\frac{2 \sum_{i,j} t_{i,j,k} \log(p(i,j,k))+1}{\sum_{i,j} t_{i,j,k}+\sum_{i,j} p(i,j,k)+1}
\end{equation}
where $k$ denotes one class among $c$ classes ($c = 3$), $t_{i,j,k}$ is equal to 1 if the pixel ($i, j$) belongs to class $k$, $p(i, j, k)$ denotes the probability of pixel ($i, j$) of being class $k$, $n$ is the number of pixels in the patch. Cross-entropy is the most popular loss function for classification tasks. Soft dice loss \cite{milletari2016v} was adapted from the Dice coefficient to calculate the similarity between two images. One was added in (3) to ensure that the function is not undefined when $t_{i,j,k} = p(i, j, k) = 0$ \cite{jadon2020survey}. Here, we associated $CE$ with $D$ for neurons and inter-cell contours to force the network to distinguish the two classes from the tissue. The weighting factors (0.5, 0.3 and 0.2) indicate the contribution of each item to the compound loss, were empirically defined constants and used to deal with class imbalance: $CE$ and the sum of $D$ (neuron and contour) had the same weight, while the weight of $D_{neuron}$ was slightly higher than that of $D_{contour}$ because the neuron class was the most important in our case. 

\subsection{Post-processing}
The most straightforward post-processing is to apply a threshold to the probability map, yet one threshold will not fit all tested neurons with varying sizes, shapes and intensities. Better segmentation can be achieved using more sophisticated methods such as graph partitioning or distance transform. However, these methods increase the computation cost considerably and introduce several hyperparameters that need to be defined empirically. In this study, we propose an efficient and generic post-processing approach, as presented in Algorithm \ref{alg1}. The output of our network is a 3-channel probability map, presented with an RGB image. Each channel corresponds to one class: channel R represents the background, channel G represents the contours between touching neurons and channel B represents neurons. First, we extracted pixels that were most likely to be neuron class ($argmax(P)=neuron$), denoting these pixels to 1 and other pixels to 0 to create a binary image. Second, we applied the ultimate erosion on the binary image with a disk-shaped structuring element, whose radius was equal to 10 pixels, the same as the radius of the smallest neuron expected \cite{o2007hippocampal}, ensuring that no more than one neuron would be erased during erosion to prevent under-segmentation. We hypothesized that the inter-cell class could separate entirely or partially the touching neurons. The second case often occurs in dense regions where the inter-cell class is not sufficient to cut touching neurons entirely. However,  it could create an initial concavity between cells, which would provide an optimal condition for performing ultimate erosion to complete the separation process. We labeled the ultimate residues, each residue representing an individualized neuron. Then, we proposed a dynamic dilation reconstruction using the same structuring element: each residue was dilated with the same number of erosion applied to produce an approximation of their original size. Due to the disk-shaped structuring element, the dilated residues might have an unnatural smooth shape. We used WS to restore the refined morphologic information: the dilated residues were used as seeds to initialize the WS expansion, constrained by a binary mask which merged neuron and inter-cell contour classes (1 if one pixel belonged to neuron or inter-cell contour, 0 otherwise). The merging of inter-cell contour and neuron channels aimed to restore the cell pixels lost due to our artificially created inter-cell class. 

The only parameter of our post-processing to be set is the size of the structuring element of ultimate erosion, which is equal to that of the smallest neuron in the dataset. It is easy to configure and generic, it can be applied independently to other DL-based nuclei segmentation methods using a similar strategy. 

\begin{algorithm}
  \DontPrintSemicolon
    
    \KwInput{A three-channel probability map}
    \KwOutput{Neuron instance segmentation}
    Create binary mask based on the neuron channel \;
    Ultimate erosion with a disk $S (r=10 pixels)$ \;
    \For{each ultimate residue $U_i$}
    {
      $N_i$ $\gets$ number of erosion before $U_i$ being removed \;
      dilation using $S$, $U_i$ $\gets$ $dil(U_i)$, $N_i$ $\gets$ $N_i-1$ \;
      repeat dilation until $N_i=0$
    }
    $M$ $\gets$ fusion of neuron channel and inter-cell channel \;
    Apply WS, markers $\gets$ $U$, mask $\gets$ $M$ \;
    \Return segmented neurons
  \caption{Post-processing using mathematical morphology}\label{alg1}
  \end{algorithm}

\subsection{Evaluation metrics}
We aim to establish a framework which consists of the neural network and post-processing. We performed comprehensive comparisons for neural networks and post-processing approaches respectively. Four tasks, including detection, instance segmentation, semantic segmentation, and counting, were evaluated using five metrics. They are the F1 score (F1-det) for detection, the relative count error (RCE) for counting, the Dice score for semantic segmentation, and the F1 score (F1-seg) and Aggregated Jaccard Index (AJI) for instance segmentation. 

The tasks of detection and counting were evaluated at the object level by matching the predicted neurons with the point annotation. A predicted neuron was considered as a true positive (TP) when it was superimposed with precisely one point annotation. Otherwise, it was defined as a false positive (FP: not superimposed with any centroid) or a false negative (FN: superimposed with more than one centroid). FN also included the case that no neuron was detected in the location of a centroid. With TP, FP and FN, we computed precision (P), recall (R) and F1-det, and RCE as follows: 
\begin{equation}
P = \frac{TP}{TP + FP}; R = \frac{TP}{TP + FN}; F1 = 2 \times \frac{P \times R}{P + R}
\end{equation}
\begin{equation}
  RCE = \frac{\left | FP-FN \right |}{TP+FN}
\end{equation}
RCE is the ratio of count error over the number of neurons identified with centroids.  

Furthermore, we estimated how well-segmented neurons matched synthetic masks at the pixel level. The $IoU$ score $\frac{\left | A\bigcap B \right |}{\left | A\bigcup B \right |}$ was computed for all pairs of objects, where $A$ is a predicted neuron and $B$ is the corresponding ground truth mask. When segmentation of a neuron covered the mask completely, the IoU score was 1. Since it is almost impossible to perform two identical segmentations, even for an expert, we selected 0.5 as the threshold of minimum $IoU$ to identify correct segmentation. In this case:  the TP was defined as the IoU greater than 0.5 between the predicted neuron and the synthetic mask. Otherwise, it was a FP or a FN. We computed the P, R and F1 score (F-seg) to evaluate the segmentation performance with this new criterion. Dice coefficient and Aggregated Jaccard Index (AJI) \cite{kumar2017dataset} were also calculated to evaluate the segmentation at the pixel level. The AJI is defined as:
\begin{equation}
AJI = \frac{\sum_{i=1}^{N}\left | G_i \bigcap S(G_i)\right |}{\sum_{i=1}^{N}\left | G_i\bigcup S(G_i) \right |+\sum_{u=1}^{U}\left | S_u \right |}
\end{equation}
Where $G_i$ is one ground truth object, $S(G_i)$ is the segmented object that maximizes the $IoU$ with $G_i$. $U$ is the set of segmented objects that have not been assigned to any ground truth object. AJI is the most stringent among all evaluation metrics. It aims to penalize errors at both object and pixel level. It would also help us to distinguish methods that score similarly on other metrics.  

\subsection{Implementation details}
The network was implemented using Tensorflow and Keras. The encoder was pre-trained on ImageNet. The learning rate started from 1e-4 and decreased gradually during the training. The model was trained for 100 epochs during 40h, with Adam optimizer. We monitored the training and validation loss of each epoch and saved the model with the lowest validation loss (at 44th epoch). Test Time Augmentations (TTA) of flipping ($\times$2) and rotation ($\times$4) were applied. This work was conducted on a workstation equipped with bi-processors (operating system: Ubuntu 16.04 LTS 64-bits, CPU: Intel Xeon gold 5218 at 2.3 GHz, RAM: 128 GB, GPU: NVIDIA Quadro RTX 5000 with 16 GB memory). 

\subsection{Compared methods and parameter settings}
We performed comparisons of both neural networks and post-processing approaches. For neural network benchmarks, we compared the proposed EfficientUNet with an unsupervised method proposed in \cite{you2019automated} and the following state-of-the-art networks, most were U-Net-like: UNet \cite{cui2019deep, ronneberger2015u}, HRNet \cite{wang2020deep}, an ensemble model \cite{wu2021evaluation} of eight encoder-decoder networks and each constitutive network. The encoders included three ResNets (34, 101, 152), two Dual Path Networks DPN-92 (with sigmoid and softmax activation, respectively), two DenseNets (121, 169) and one Inception-ResNet. It is worth noting that we named the eight encoder-decoder networks after their encoder for simplicity.

The method \cite{you2019automated} was unsupervised and did not require any additional training. All DL methods were trained on the same dataset with synthetic annotations. UNet was trained from scratch with the same configuration described in \cite{cui2019deep}. HRNet was trained from scratch, the training details were the same as EfficientUNet, see 2.7. The eight constituents of the ensemble model were trained separately with encoders pre-trained on ImageNet using the configuration described in \cite{caicedo2019nucleus}. In this part, we applied the same post-processing \cite{caicedo2019nucleus} to eliminate the effect of different post-processing for DL methods. It introduced a second-stage training procedure, training a regression model with neuron morphological information to predict the IoU. We created three groups of cell candidates using three thresholds (0.6, 0.7, and 0.8), and only the candidate with the highest predicted IoU was retrained for each object. The segmentation would be removed if the predicted IoU was too low, in order to reduce the number of FPs.

Another contribution of this work is the post-processing scheme. Ablation experiments were carried out to demonstrate the benefits of the proposed post-processing. We compared the proposed post-processing scheme to four classical methods: the first, which is the most often used method, includes applying WS to the probability map's thresholded ($\geq$ 0.5) cell channel, which served as a baseline for this section.  The second is the winning method of \cite{caicedo2019nucleus}, which was the same approach that we used for the comparison of networks. Additionally, we assessed the post-processing scheme of \cite{zeng2019ric}, which refines the neuron class by removing ambiguous pixels that are likely to be the contour class, followed by a dilation process to recover the neuron form. The last is a widely-used distance-transformation-based technique \cite{naylor2018segmentation, xie2020integrating}. Distance transformation is applied to the neuron class to obtain the distance map. With a minimum allowed distance being 20 pixels (minimum diameter of neurons), the local maximums of the distance map are further used as the seeds to initialize WS. Only the ultimate erosion and the dynamic dilation in Algorithm \ref{alg1} are replaced by the distance transformation to fairly compare the separation capacity of the two techniques. All approaches were applied on the same probability maps, the prediction results of EfficentUNet.

\section{Results}
In this section, we first show the representativeness of the test set versus the training set, see 3.1. In 3.2, we quantitatively evaluate the quality of the synthetic mask by comparing it with three manual segmentations. In 3.3, we present a comprehensive comparison of the proposed EfficientUNet and several state-of-the-art methods. In 3.4, we illustrate the advantage of our post-processing strategy in comparison with other classical techniques. 

\subsection{Qualitative evaluation of datasets composition}
\begin{figure*}
\centering
\includegraphics[width=0.9\linewidth]{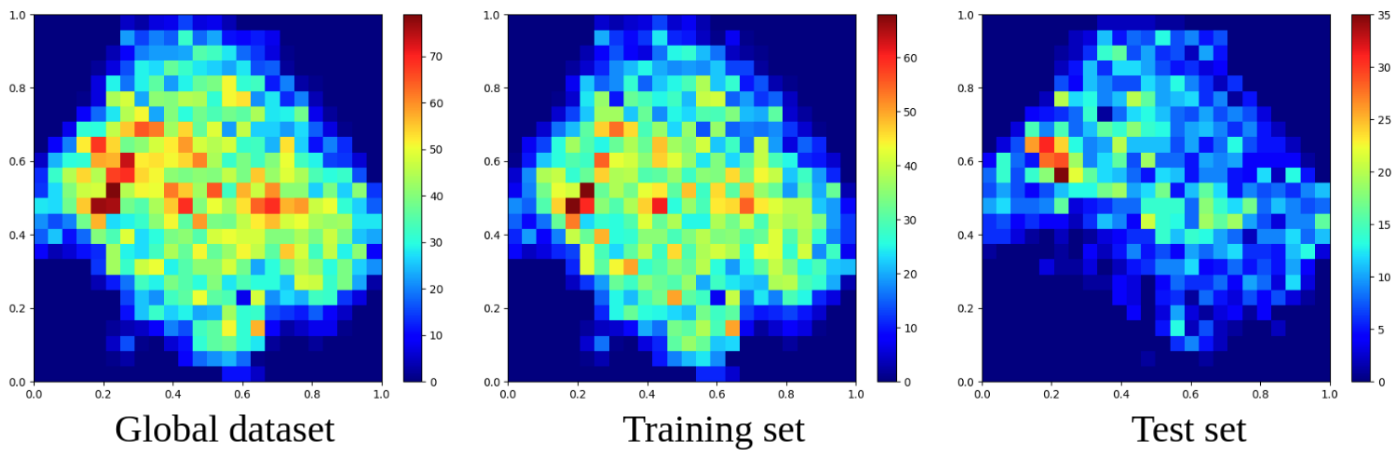}
\caption{Heatmaps of t-SNE projection. Left to right: all data, training data and test data.}
\label{fig4}
\end{figure*}

As shown in Fig. \ref{fig4}, we plotted the heatmaps (number of bins = 25) based on the t-SNE projection of the whole dataset, the training set and the test set, respectively. X and Y were normalized to [0,1], representing the two dimensions of the embedded space. Since test data have similar feature distribution (t-SNE plots of two distributions present overlap) compared to that of training data, we can consider the test set as representative of the whole dataset.

\subsection{Quantitative validation of synthetic masks}
\begin{figure*}
\centering
\includegraphics[width=0.7\linewidth]{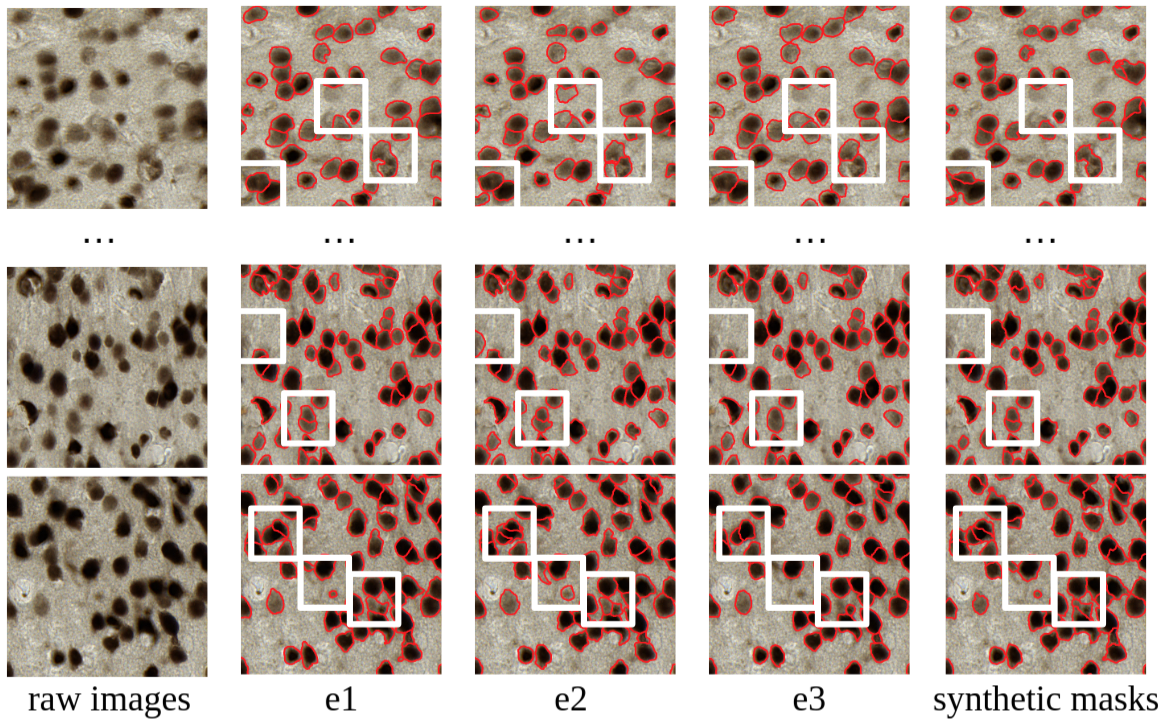}
\caption{Comparison of manual segmentations and synthetic masks. The white squares show the inconsistencies between manual segmentations.}
\label{fig5}
\end{figure*}

The objective of our mask synthesis pipeline was to minimize manual effort, the expert labeled only the centroid of neurons instead of identifying the contour of neurons. We considered the point annotations unbiased, and evaluated the binary segmentation as well as the region growing process.  Table \ref{tab1} reports the average IoU score (mIoU) and Dice coefficient between the synthetic masks and three manual annotations (denoted as e1, e2 and e3) respectively. We also computed the scores between manual annotations to show the inter-expert variability. We observed that the evaluation scores varied when we compared the synthetic masks to different manual annotations. Overall, our synthetic masks were of good quality, in particular, with the e1 group as the reference. Fig. \ref{fig5} shows inconsistent manual segmentations between experts in three of five images. On average, the scores of synthetic annotations were at the same level of the inter-experts scores, with a tiny difference on Dice (-0.6\%) and a slight decrease (-2.4\%) of mIoU. On the other hand, the experts spent 2.5 hours on annotating five small images (500 $\times$ 500 pixels). The entire training dataset (11k images of 224 $\times$ 224 pixels) would take over two months to annotate manually, demonstrating the need of the semi-automatic mask synthesis pipeline. 

\begin{table*}
\centering
\caption{Average IoU score (mIoU) and Dice coefficient between the synthetic masks and three manual annotations (denoted as e1, e2 and e3) respectively.}
\label{tab1}
\begin{tabular}{lllllllll}
  \toprule
  \multirow{2}{*}{Evaluation metrics} & \multicolumn{4}{c}{Synthetic vs experts} & \multicolumn{4}{c}{Inter-experts} \\
  \cmidrule(lr){2-5} \cmidrule(lr){6-9}
  & ours vs e1 & ours vs e2 & ours vs e3 & mean & e1 vs e2 & e1 vs e3 & e2 vs e3 & mean \\
  \hline
  mIoU & 0.774 & 0.706 & 0.724 & 0.735 & 0.753 & 0.769 & 0.755 & 0.759\\
  Dice & 0.914 & 0.888 & 0.888 & 0.897 & 0.898 & 0.91 & 0.9 & 0.903 \\
  \bottomrule
\end{tabular}
\end{table*}

\subsection{Comparison of segmentation methods}
Table \ref{tab2} reports the performance of the tasks of detection (F1-det), counting (RCE), instance segmentation (F1-seg and AJI) and semantic segmentation (Dice), as well as the computational complexity (\#Params: number of trainable parameters and number of FLOPs: Floating-Point Operations) of the proposed EfficientUNet and other 12 methods as mentioned in section 2.8. We did not compute the Dice for the unsupervised method \cite{you2019automated}, because it used the same binary RF segmentation that constrained our synthetic masks. All DL methods except UNet had a very high F1-det (> 0.92) with low variance. The ensemble model outperformed the others on the tasks of detection (F1-det: 0.931) and counting (RCE: 0.026). In contrast, the best model for the tasks of semantic and instance segmentation was one constitutive model, DPN-soft, with the highest scores of Dice (0.951), F1-seg (0.907) and AJI (0.759). The proposed EfficientUNet was the second best model for segmentation tasks and it also performed well on the tasks of detection and counting, with similar scores to the best model for each metric (F1-det: -0.007, RCE: +0.018, Dice: -0.017, F1-seg: -0.006, AJI: -0.017). Followed by two constitutive ResNet backbone networks: ResNet-34 had good performance in the tasks of detection and semantic segmentation, while the performance declined on the task of counting. ResNet-101 was the best single model at the object level (F1-det and RCE), but it performed poorly on segmentation tasks. The scores of HRNet were comparable to those of EfficientUNet, but the scores on segmentation tasks were slightly lower (Dice, F1-seg and AJI). The scores of UNet were the lowest in all the tasks, at both object and pixel levels. The unsupervised method \cite{you2019automated} achieved good results on segmentation tasks: 0.87 in F1-seg and 0.729 in AJI, while it performed worse on the tasks of detection and counting. Further considering the computational complexity of the DL models, the proposed EfficientUNet was the best among all single models. The FLOPs of the proposed EfficientUNet (15.13 G) were similar to that of the lightest model (UNet, FLOPs: 15.29 G), regardless its model size was six times that of UNet. By contrast, although the performance of DPN-soft was slightly better than that of EfficientUNet, the complex structure (Params: 133.5 M, FLOPs: 51.64 G) makes it complicated to be trained and applied efficiently to larger-scale images.  

\begin{table*}
\centering
\caption{Comparison of neuron detection, counting, instance segmentation and computational complexity of different automatic methods. Best and second best results are in bold with the best also underlined. FLOPs are computed for a 224 $\times$ 224 input RGB image.}
\label{tab2}
\begin{tabular}{llllllll}
  \toprule
  \multirow{2}{*}{Model} & \multicolumn{7}{c}{(Mean\( \pm \) Standard deviation)} \\
  \cmidrule(lr){2-8}
  & \#Params(M) & FLOPs(G) & F1-det & RCE & Dice & F1-seg & AJI \\
  \hline
  You et al. & - & - & 0.889\( \pm \)0.019 & 0.076\( \pm \)0.062 & - & 0.87\( \pm \)0.017 & 0.729\( \pm \)0.03 \\
  EfficientUNet & 39.25 & 15.29 & 0.924\( \pm \)0.012 & 0.044\( \pm \)0.033 & \textbf{0.934\( \pm \)0.024} & \textbf{0.901\( \pm \)0.018} & \textbf{0.742\( \pm \)0.048} \\
  U-Net & 5.75 & 15.13 & 0.819\( \pm \)0.061 & 0.264\( \pm \)0.099 & 0.890\( \pm \)0.017 & 0.715\( \pm \)0.087 & 0.473\( \pm \)0.137 \\
  HRNet & 9.5 & 26.55 & 0.924\( \pm \)0.011 & 0.044\( \pm \)0.041 & 0.933\( \pm \)0.019 & 0.895\( \pm \)0.023 & 0.732\( \pm \)0.049 \\
  \hline
  DenseNet-121 & 17 & 30.26 & 0.920\( \pm \)0.018 & 0.041\( \pm \)0.033 & 0.830\( \pm \)0.047 & 0.801\( \pm \)0.09 & 0.615\( \pm \)0.053 \\
  DenseNet-169 & 18.5 & 44.26 & 0.924\( \pm \)0.012 & 0.043\( \pm \)0.041 & 0.883\( \pm \)0.032 & 0.863\( \pm \)0.034 & 0.682\( \pm \)0.064 \\
  ResNet-34 & 23.75 & 7.81 & \textbf{0.929\( \pm \)0.014} & 0.067\( \pm \)0.034 & \textbf{0.934\( \pm \)0.015} & \textbf{0.901\( \pm \)0.024} & 0.729\( \pm \)0.047 \\
  ResNet-101 & 47.25 & 52.08 & \textbf{0.929\( \pm \)0.013} & \textbf{0.036\( \pm \)0.034} & 0.911\( \pm \)0.023 & 0.875\( \pm \)0.034 & 0.690\( \pm \)0.049 \\
  ResNet-152 & 62.25 & 58.99 & 0.926\( \pm \)0.013 & 0.039\( \pm \)0.039 & 0.918\( \pm \)0.018 & 0.878\( \pm \)0.026 & 0.704\( \pm \)0.038 \\
  Incep-ResNet & 63.75 & 40.98 & 0.925\( \pm \)0.014 & 0.081\( \pm \)0.043 & 0.923\( \pm \)0.026 & 0.894\( \pm \)0.022 & 0.729\( \pm \)0.053 \\
  DPN-sig & 133.5 & 51.64 & 0.921\( \pm \)0.016 & 0.053\( \pm \)0.03 & 0.910\( \pm \)0.029 & 0.864\( \pm \)0.05 & 0.679\( \pm \)0.064 \\
  DPN-soft & 133.5 & 51.64 & 0.928\( \pm \)0.011 & 0.043\( \pm \)0.038 & \underline{\textbf{0.951\( \pm \)0.011}} & \underline{\textbf{0.907\( \pm \)0.022}} & \underline{\textbf{0.759\( \pm \)0.048}} \\
  \hline
  Ensemble & - & - & \underline{\textbf{0.931\( \pm \)0.011}} & \underline{\textbf{0.026\( \pm \)0.023}} & 0.918\( \pm \)0.026 & 0.88\( \pm \)0.036 & 0.709\( \pm \)0.046 \\ 
  
\bottomrule
\end{tabular}
\end{table*}

\begin{figure*}
\centering
\includegraphics[width=0.9\linewidth]{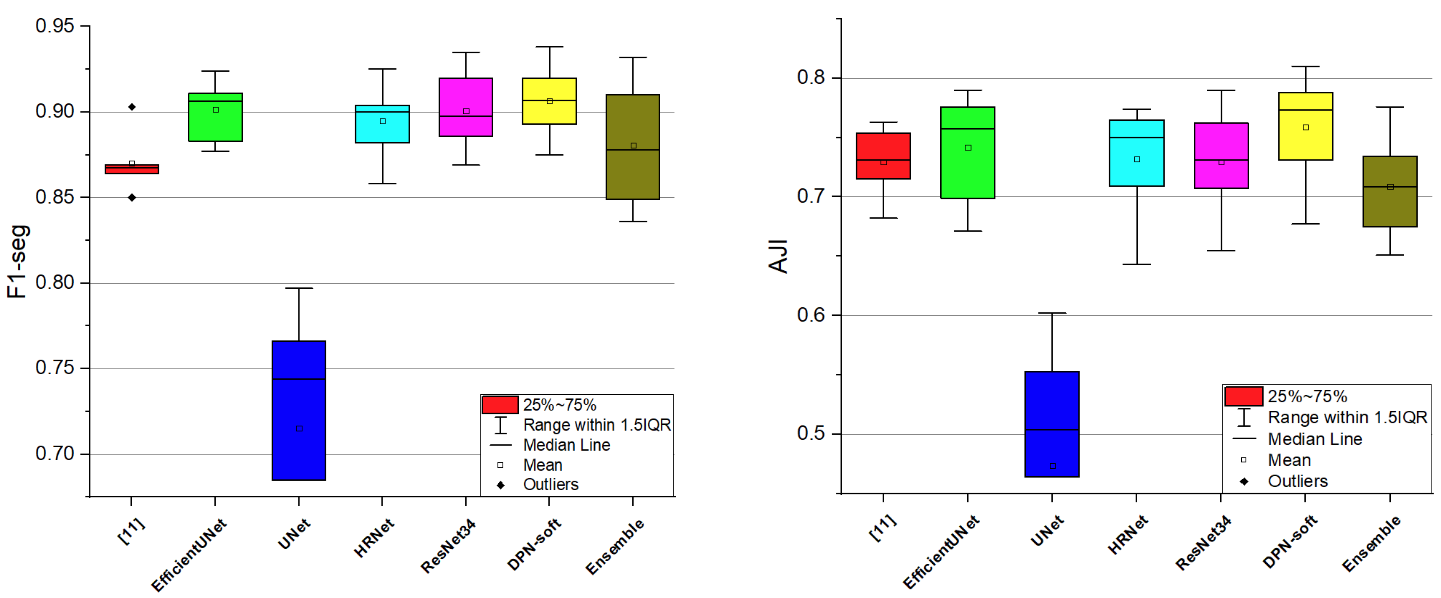}
\caption{Boxplot of F1-seg and AJI for different methods for comparative analysis. Left panel: F1 scores of segmentation. Right panel: AJI scores(outlier values of UNet are below the vertical scale range).}
\label{fig6}
\end{figure*}

As shown in Table \ref{tab2}, most methods reach a high F1-det and Dice, which is not helpful for performance comparison. On the other hand, F1-seg and AJI provide distinct differences between methods, revealing performance at both object and pixel levels. Fig. \ref{fig6} compares F1-seg and AJI of the unsupervised method You et al. \cite{you2019automated}, the proposed EfficientUNet, UNet, HRNet, the two best constitutive models (ResNet-34 and DPN-soft) and the ensemble model. What stands out in the figure is that UNet performed poorly on the task of instance segmentation. In terms of F1-seg, both the proposed EfficientUNet and DPN-soft provided good and robust performance. In particular, the difference in EfficientUNet scores across regions was less than 4\%, which indicates that it performed well on all anatomical regions tested. Although the median score of HRNet was similar to that of ResNet-34, the interquartile range was slightly lower than that of ResNet-34. The mean score of the unsupervised method You et al. \cite{you2019automated} was at the same level as that of the ensemble model, but it was more robust, with the slightest variation between regions. AJI illustrates differences between methods in a more significant way, the scores of all the methods were decreased. DPN-soft and the proposed EfficientUNet remained the two best models. As well as HRNet and ResNet-34, their scores were similar but slightly lower than the best models. Surprisingly, You et al. \cite{you2019automated} was the second-last method of F1-seg, showing better performance than the ensemble model with AJI: it reached the same level as HRNet and ResNet-34. 

\begin{figure*}
\centering
\includegraphics[width=1\linewidth]{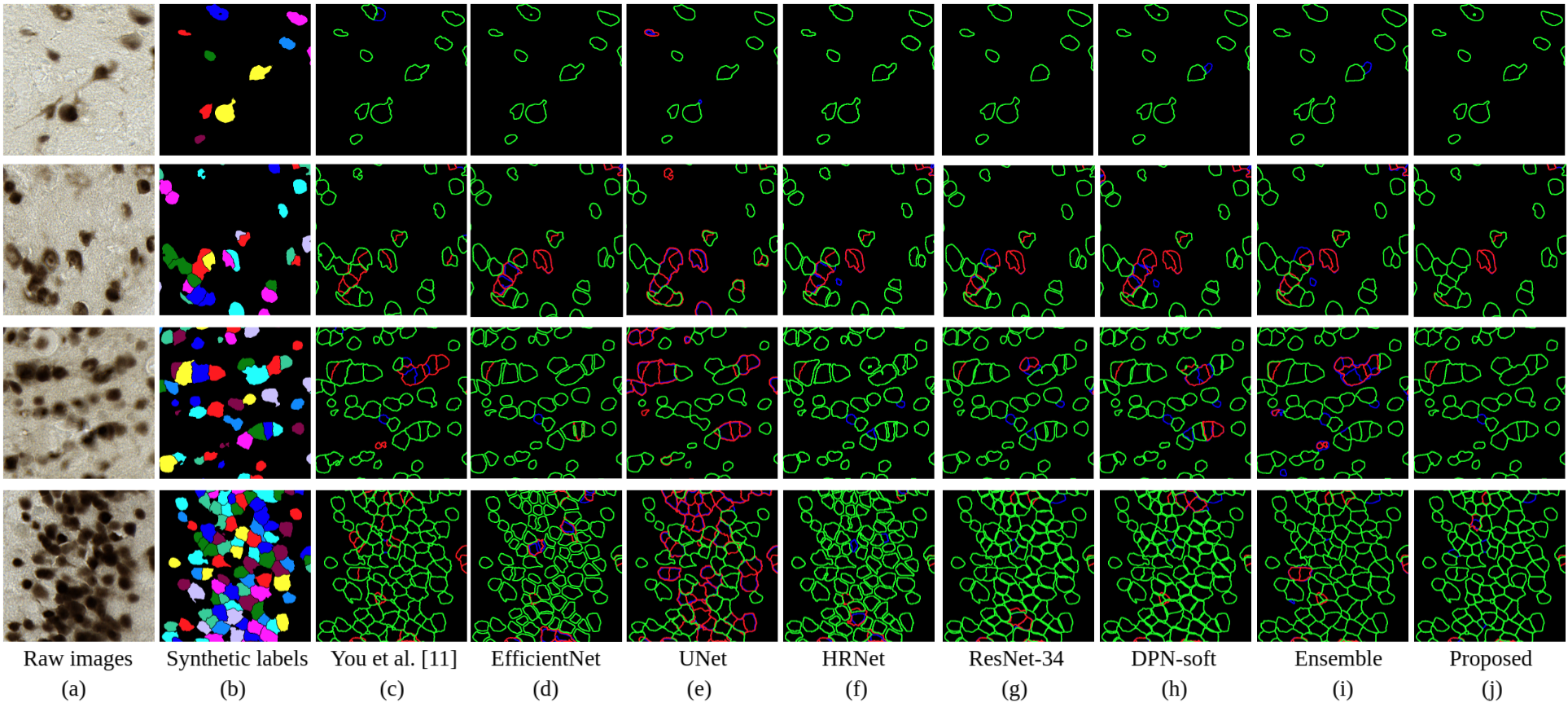}
\caption{Original images and segmentation results of our method and other approaches in different regions with different neuron density. (a) raw images, (b) synthetic labels, (c-j) results of You et al. \cite{you2019automated}, EfficientUNet, UNet, HRNet, ResNet-34, DPN-soft, the ensemble model and the entire proposed framework. Top to Down: (1) thalamus, (2) caudate, (3) cortex and (4) hippocampus. The green contours represent segmentations that overlap the synthetic ground truth (IoU > 0.5), the blue contours represent over-segmentations and the red contours represent missing and under-segmentations}
\label{fig7}
\end{figure*}

Fig. \ref{fig7} shows comparative results of representative patches from different anatomical regions. Besides the methods mentioned above (in Table \ref{tab2} and Fig. \ref{fig6}), we also report the segmentation results of the complete proposed pipeline (network and post-processing) to visualize the improvement brought by the proposed post-processing. If the IoU between a segmented object and the ground truth is superior to 0.5, we highlighted the contours in green. Otherwise, the contours are displayed either in blue for over-segmentation or red for under-segmentation and missing detection respectively. You et al. \cite{you2019automated} (column c) segmented most neurons correctly in sparse regions, while under-segmentation often occurred when the neurons aggregate. UNet (column e) suffered from under- and missing segmentation, but it was also ineffective in preserving the shape of neurons. The segmentations of EfficientUNet (column d), HRNet (column f), ResNet-34 (column g) were roughly comparable, performing well in all anatomical regions, from the thalamus (sparse) to the hippocampus (dense). Especially EfficientUNet caused fewer segmentation errors in caudate and cortex. DPN-soft (column h) and the ensemble model (column i) caused over-segmentations in thalamus and caudate, and under-segmentations in cortex and hippocampus. EfficientUNet with the proposed post-processing (column j) made fewer segmentation errors than the others in any anatomical region, suggesting that the proposed post-processing can successfully correct segmentation errors of the neural network (see columns d and j).   

\subsection{Comparison of post-processing schemes}
Table \ref{tab3} and Table \ref{tab4} compare the performance of the instance segmentation task (F1-seg and AJI) and the processing time of different post-processing techniques respectively. Although the probability maps were the same (EfficientUNet), the results vary considerably between the different post-processing methods. The proposed method appeared to be the best for all anatomical regions, with an average F1-seg of 0.917 and AJI of 0.774. The distance transformation achieved comparable scores with slight decreases, and it was less computationally expensive than the proposed method. However, it over-segmented circle-shaped neurons (see Supplementary Fig. 5). Compared to the baseline,  the regression model of Topcoders \cite{caicedo2019nucleus} brought a slight improvement (2\% in both F1-seg and AJI), whereas it took 460.6 s to process one image, which is 27 times that of the baseline. Zeng et al. \cite{zeng2019ric} was the fastest method, and it required only 7.88s to process one image. However, it was also the method that performed the worst, with a decline of 9.2\% and 14\% in F1-seg and AJI compared to the proposed method. Among all tested anatomical regions, higher scores were found in the region with high contrast (subiculum) and in the region where only a few neurons were aggregated (caudate). In contrast, the scores dropped in the region with high neuron density (hippocampus) and the region with low contrast with the background (thalamus). By comparing the proposed method and the baseline, ours significantly improved the instance segmentation, with an average gain of 4\% and 4.6\% in F1-seg and AJI, respectively. In particular, for dense regions like the cortex and the hippocampus, where lots of neurons aggregate, the proposed method increased AJI by 7.7\% and 6.3\%, respectively.

\begin{table*}
\centering
\caption{Instance segmentation performance using different post-processing techniques. The best results are in bold.}
\label{tab3}
\begin{tabular}{lllllllllll}
  \toprule
  \multirow{2}{*}{Region} & \multicolumn{5}{c}{F1-seg} & \multicolumn{5}{c}{AJI} \\
  \cmidrule(lr){2-6} \cmidrule(lr){7-11}
  & baseline & Topcoders & Zeng et al. & Dist & Ours & baseline & Topcoders & Zeng et al. & Dist & Ours \\
  \hline
  caudate & 0.885 & 0.903 & 0.82 & 0.906 & \textbf{0.916} & 0.754 & 0.775 & 0.641 & 0.787 & \textbf{0.79} \\
  cortex & 0.87 & 0.896 & 0.822 & 0.909 & \textbf{0.915} & 0.689 & 0.72 & 0.598 & 0.759 & \textbf{0.766} \\
  hippocamp & 0.857 & 0.877 & 0.828 & 0.899 & \textbf{0.902} & 0.661 & 0.671 & 0.614 & 0.722 & \textbf{0.724} \\
  subiculum & 0.912 & 0.924 & 0.877 & 0.939 & \textbf{0.94} & 0.783 & 0.79 & 0.69 & 0.819 & \textbf{0.821} \\
  thalamus & 0.876 & 0.911 & 0.781 & 0.896 & \textbf{0.916} & 0.76 & 0.776 & 0.662 & 0.758 & \textbf{0.777} \\
  overall & 0.878 & 0.901 & 0.825 & 0.91 & \textbf{0.917} & 0.723 & 0.742 & 0.634 & 0.767 & \textbf{0.774} \\
  \bottomrule
\end{tabular}
\end{table*}

\begin{table}
\centering
\caption{Processing time for one image of 5k $\times$ 5k using different post-processing approaches.}
\label{tab4}
\resizebox{\columnwidth}{!}{%
\begin{tabular}{llllll}
  \toprule
  Method & baseline & Topcoders & Zeng et al. & Dist & Ours \\
  \midrule
  Time(s) & 17.17 & 460.6 & 7.88 & 28.3 & 118.57 \\
  \bottomrule
\end{tabular}%
}
\end{table}

\begin{figure*}
\centering
\includegraphics[width=0.9\linewidth]{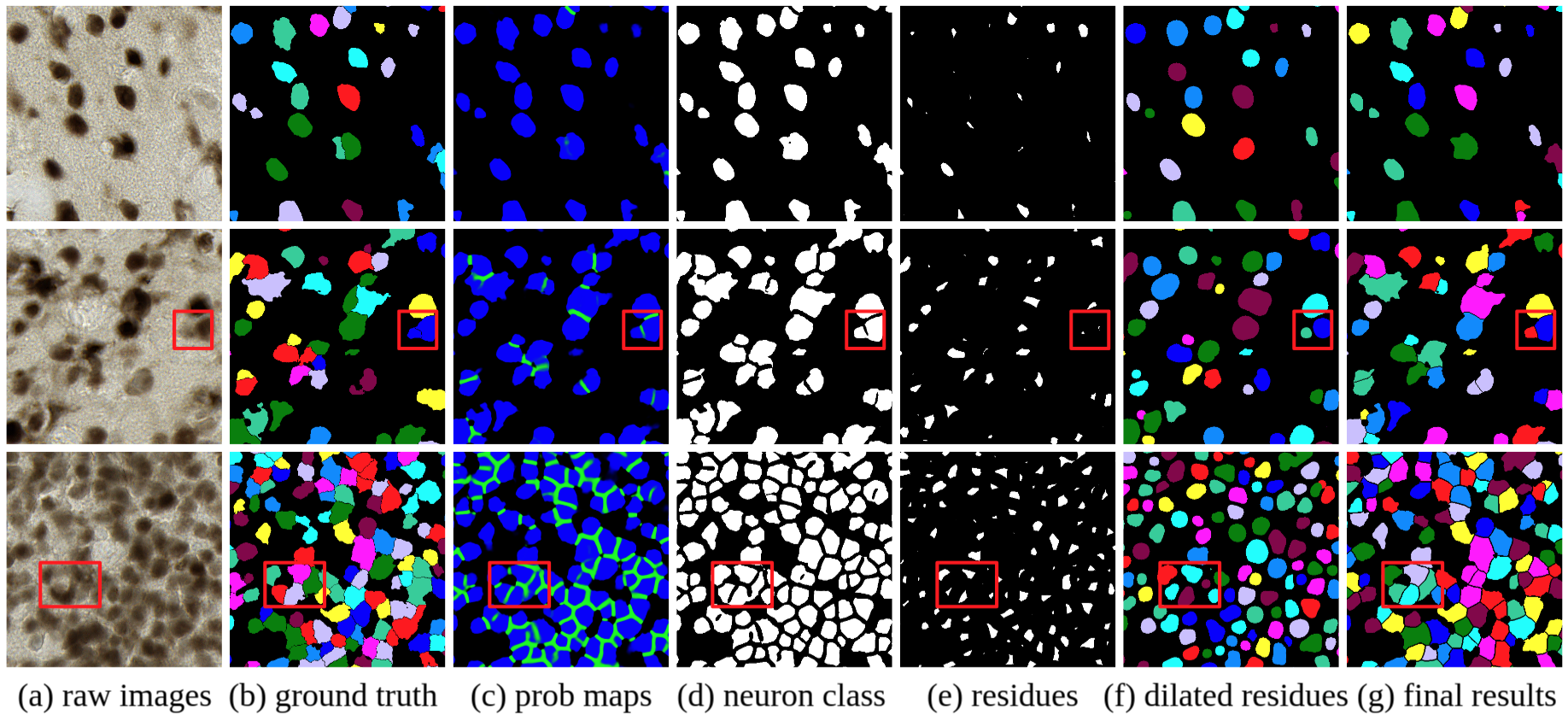}
\caption{Intermediate results of the proposed post-processing. Top to down: (1) caudate, (2) cortex and (3) hippocampus. Left to right: (a) original images, (b) ground truth, (c) probability map of deep network, (d) binary mask of neuron channel, (e) ultimate residues, (f) reconstructed residues and (g) final segmentation after WS. The red square highlights the neurons that the neural network failed to separate but being fully segmented through the proposed post-processing.}
\label{fig8}
\end{figure*}

Fig. \ref{fig8} presents the results for each intermediate step of our post-processing scheme. The inter-cell class allowed our neural network to segment the sparse region like caudate and most neurons in the cortex. However, our network sometimes failed to separate touching neurons, as illustrated by the red square in Fig. \ref{fig8} (b, c). This problem became troublesome for the hippocampus region, where many neurons aggregated, which would cause the under-estimation in neuron population counting. In this particular case, the ultimate erosion process in our scheme could bring a critical advantage. Although the inter-cell class did not completely separate the touching neurons, it provided an optimal condition to apply further erosion: an initial concavity. Fig. \ref{fig8} (d) shows the ultimate residues computed. We observed that the neurons that the neural network had not separated were now fully individualized. By applying the same iteration number of dilation with the same structuring element as erosion on each ultimate residue, we restored the coarse morphological information of neurons, as illustrated in Fig. \ref{fig8} (e). Finally, the refined segmentation was obtained with the WS using dilated residues as seeds. A combination of inter-cell and neuron classes was used to constrain the expansion, allowing to eliminate the artificial gap between touching cells by reassigning the inter-cell pixels to neurons.

\subsection{Assessment of the generalizability of the proposed method}
Table \ref{tab5} reports the object-level segmentation of the proposed method on the supplementary test set, including cortex images of two macaques, one microcebus and one mouse. The purpose of this experiment was to evaluate the generalizability of the proposed method on data different from the training set. The results on cortex images of the initial test set are illustrated as a reference. For the task of detection, the scores decreased on other animal subjects, but no significant differences were observed, with the largest decrease being 2.23\% that was found in the mouse. As for the task of counting, the method made more errors on other animal subjects except for the microcebus, which had the lowest RCE of 0.03\%. The RCE of the macaques and the mouse was approximately on the same level, with an increase of 5\% compared to the reference. 

\begin{table*}
\centering
\caption{Object-level segmentation performance of the proposed method on the supplementary test set.}
\label{tab5}
  \begin{tabular}{lllllll}
  \toprule
  data supp & ref macaque & macaque 1 &  macaque 2 & microcebus & mouse & mean \\
  \midrule
  F1-det & 0.93 & 0.918 & 0.927 & 0.921 & 0.908 & 0.918 \\
  RCE & 0.02 & 0.075 & 0.073 & 0.003 & 0.071 & 0.056 \\
  \bottomrule
\end{tabular}
\end{table*}

\section{Discussion}
Unbiased quantification of individualized cells is essential for many biomedical analyses. One challenging application field is counting and individualizing the neuronal cells as their size, shape and density vary from one anatomical region to another. Recent studies demonstrated the importance of neuron morphology and distribution in studying cerebral functions and neurodegenerative diseases \cite{karlsen2011total, hughes1987morphometric, thu2010cell, vicar2020quantitative}. Therefore, an automatic neuron segmentation method is a cornerstone for such research. In this work, we present an end-to-end framework aiming to improve neuron detection and instance segmentation in the major anatomical regions of the macaque brain. Our mask synthesis pipeline based on pin-pointed centroids and RF binary segmentation allowed us to generate large amounts of pixel-level annotations for training, which would have been impossible to achieve with manual cell segmentation. In addition, embedding the state-of-the-art CNN (EfficientNet-B5) into a UNet-like architecture increased segmentation accuracy. Although mathematical morphology techniques such as ultimate erosion for segmenting connected components \cite{vincent1989morphological} were proposed decades ago, it requires a strong concavity prerequisite between the connected components to obtain good results. This condition was satisfied by adding the inter-cell class to the probability map. Furthermore, we proposed dynamic reconstruction as a complementary step with the ultimate erosion to further improve the morphological reconstruction.

Table \ref{tab1} and Fig. \ref{fig5} showed that pixel level annotations produced by our pipeline were comparable to those of experts, which confirmed the good quality of the synthetic annotation. After an exhaustive comparison, EfficientUNet was chosen as it was the model with the best trade-off between accuracy and computational cost. The ensemble model was the best model for the tasks of detection and counting, but it was difficult to train and maintain because it consisted of eight independent networks. As for the tasks of instance and semantic segmentation, the best model was DPN-soft. It was the heaviest individual model, with the model size and FLOPs 3.4 times that of EfficientUNet, resulting in a considerably higher inference latency. The proposed EfficientUNet was not computationally expensive but outperformed most networks with a slight decrease over the best models. An efficient network is indispensable to extend this work to a larger scale image (entire brain section, even brains). Under this constraint, EfficientUNet seems to be the best choice among all tested networks. ResNet-34 and HRNet also showed a good trade-off between accuracy and efficiency, they could be potential candidates for developing an efficient segmentation network with a slight loss of accuracy. However, the models that reported good performance (HRNet, ResNet-34, EfficientNet and DPN-soft) caused a gap  between the touching neurons, especially in the dense regions such as the hippocampus, as shown in Fig. \ref{fig7} d, f-h. One explanation is that all three classes were generally well classified with these methods, including the inter-cell class. However, the applied post-processing scheme proposed in \cite{caicedo2019nucleus} used only the probability map of neuron class to constrain the WS expansion. Our post-processing results (last column of Fig. \ref{fig7}) demonstrated that this contradiction could be eliminated by merging the neuron and the inter-cell classes as the WS mask. The results of EfficientUNet and the constitutive models suggested the superiority of UNet-like architecture on segmentation tasks. However, the basic UNet performed poorly in almost all regions because the depth and width of UNet were probably not sufficient to capture complex features of neurons.  

In previous literature, researchers concentrated on enhancing network designs to improve cell segmentation, while the post-processing phase was usually under-investigated. This work highlights the significance of this process. Despite the fact that all post-processing techniques were applied to the same probability map, the proposed method outperformed \cite{zeng2019ric} by 14\% in AJI (see Table \ref{tab3}), which is more considerable than the majority of the neural network performance differences in Table \ref{tab2}. Distance transformation was less computationally expensive than ultimate erosion, and it achieved comparative results. However, it is not suited for neurons since it cannot correctly segment circle-shaped objects. The post-processing approach of \cite{caicedo2019nucleus} slightly improved the segmentation by using a better threshold and reducing FP segmentations. However, in addition to the extra training time of the regression model, it was also computationally expensive. For each object on the image, we needed to calculate the morphological information for three cell candidates and retain only the candidate with the highest predicted IoU. On the contrary, the most time-consuming step in the proposed post-processing method, the ultimate erosion, was applied at the image scale. It considerably improved accuracy while requiring less computing time than \cite{caicedo2019nucleus}. Moreover, it can be easily applied to other DL based methods and other nuclei data without expertise-demanding parameter settings. The only parameter that needs to be adjusted is the size of the structuring element used for morphological operations, which corresponds to the size of the smallest cell. Nevertheless, one prerequisite of  the proposed method is that the cell contours need to be roughly smooth and without obvious concavity, as is the case with NeuN-stained neurons. Assume the cells have complex shapes with branches (e.g. microglia with Iba1 and astrocytes with GFAP). In that case, in addition to retraining the deep model, modifications in post-processing will be required to preserve the particular morphological information. 

A heavy data augmentation has been applied to increase the robustness of the neural network. Table \ref{tab5} illustrates the object-level segmentation results of the proposed method on various animal subjects. Our findings suggest that the proposed method remained effective despite the difference between samples and species. The performance of detection slightly decreased in macaque and microcebus images which have distinct colors with the training set. On the other hand, a significant decline was observed in the mouse image, indicating that the light stain intensity may be a more crucial challenge than the color inconsistency between the test and training images. Hence, it could be conceivably hypothesized that stain intensity augmentation is required to further enhance the model robustness. Nevertheless, as a preliminary study, this dataset contained only four images of the cortex. Further work on larger datasets with other anatomical regions needs to be investigated to confirm this observation.  

Taken together, our framework is competitive on both the tasks of detection and instance segmentation compared to other reference approaches. From point annotations to pixel-level neuron individualization, it performs well on all tested anatomical regions, with efficient architecture and more accessible parameter settings.

\section{Conclusion}
In this paper, we present an end-to-end DL framework to perform neuron detection and instance segmentation. A major problem with neural networks is the difficulty of producing precise annotations at large scales. We proposed a mask-synthesis pipeline to generate pixel-level labels using only point annotations, which considerably reduced the manual labeling effort and processing duration. This pipeline was applied to automatically generate annotations at large scale for the NeuN dataset, and the same strategy can be applied to other datasets. The efficiency of UNet-like design on segmentation was proved by a thorough comparison of networks based on the synthetic annotation. The proposed EfficientUNet, in particular, offered the optimum trade-off between accuracy and computation cost. It is, therefore, possible to be applied to large-scale biological studies. Using the probability maps of EfficientUNet, we compared various post-processing approaches and demonstrated the significance of this step on instance segmentation. The segmentation of the neural network was further enhanced by our post-processing method through ultimate erosion and dynamic reconstruction. In particular, the excellent performance in the cortex and the hippocampus enables us to envision further investigation related to brain functions and neurodegenerative diseases, for instance, quantitative assessment of neuronal loss to characterize animal models and to evaluate drug efficacy. More importantly, the proposed post-processing does not require ad-hoc parameter setting, which can be of great value in practice for non-expert users. The preliminary study on other animal subjects demonstrated the good generalizability of our framework. The decline in the mouse image demonstrated the impact of the staining intensity on the segmentation. Future work should consider integrating intensity changes during the data augmentation to increase the robustness of the model toward intensity inconsistencies. The current evaluation focused on images of 5k $\times$ 5k pixels. Patch extraction, patch prediction, and stitching were all intermediate processes that caused additional memory costs. Further works on partial image reading and writing and developments for high-performance computing are required to expand this research to whole histological sections and potentially entire brains. It would help us to better understand brain development and aging, and would also provide efficient tools to develop and validate new therapies.

\section{Declaration of Competing Interests}
The authors declare that they have no known competing financial interests or personal relationships that could have appeared to influence the work reported in this paper. 

\section{Acknowledgements}
The authors wish to thank the anonymous reviewers for their insightful comments and suggestions. This work was supported by DIM ELICIT grants from Région Ile-de-France. The authors would like to acknowledge Camille Mabillon for manual segmentation, Zhenzhen You for point annotation and Géraldine Liot, Suzanne Lam, Fanny Petit and Marc Dhenain for providing supplementary mouse and microcebus brain images. 



\bibliographystyle{model1-num-names}
\bibliography{refs}

\includepdf[pages=-]{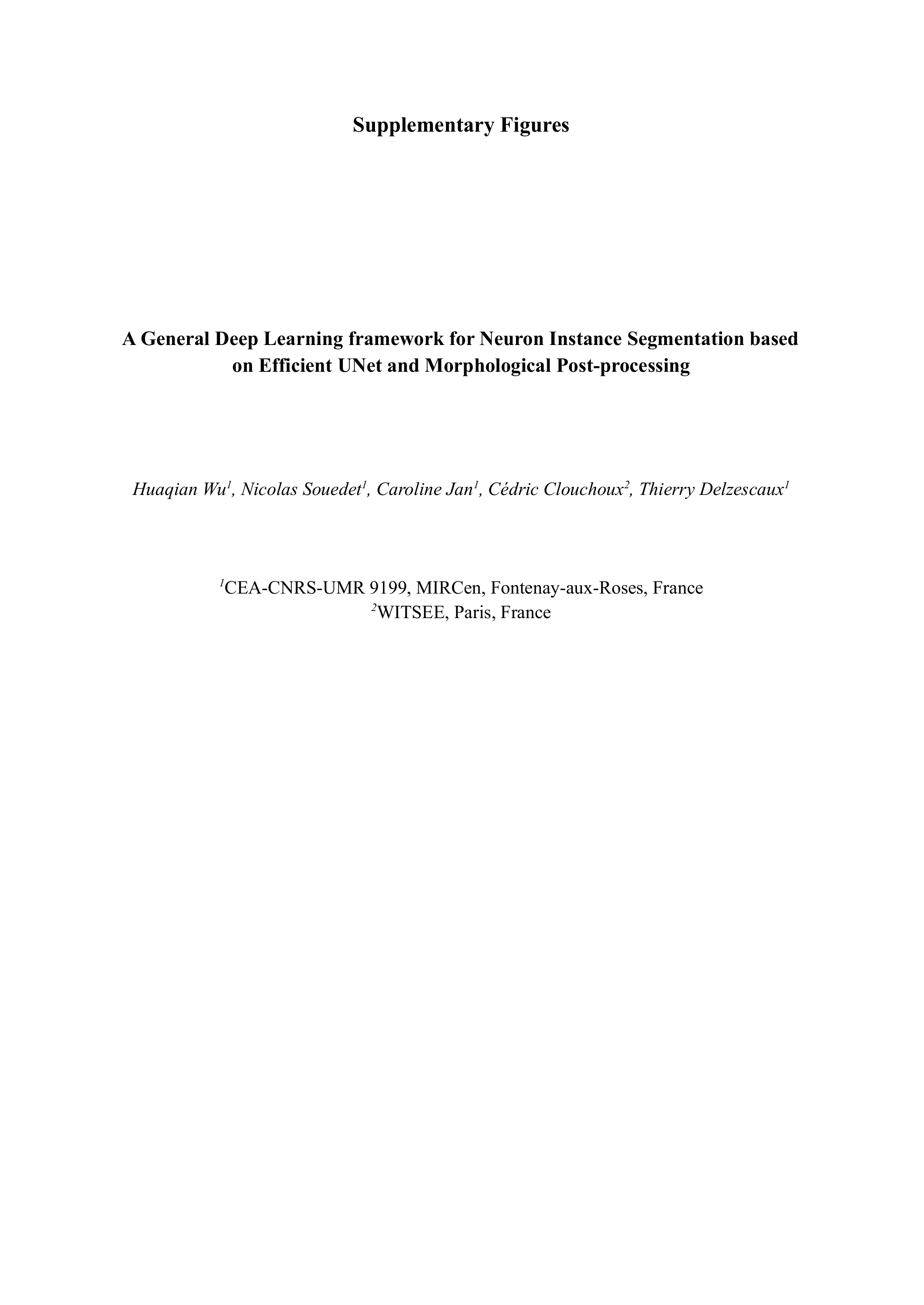}

\end{document}